\documentclass[aps,prd,a4paper,10pt,amsfonts,amssymb,nofootinbib,onecolumn,longbibliography]{revtex4}
\usepackage[utf8]{inputenc}
\usepackage[a4paper,
            bindingoffset=0.2in,
            left=1.5cm,
            right=1.5cm,
            top=1in,
            bottom=1.5cm,
            footskip=.25in]{geometry}
\usepackage{graphicx}
\usepackage{subfig}
\usepackage{caption}
\usepackage{float}
\usepackage{ulem}
\usepackage{listings}
\usepackage{xcolor}
\usepackage{upgreek}
\usepackage{amsmath}
\usepackage{geometry}
\begin{document}
\title{Baryon anticorrelations and the Pauli principle in {\tt PYTHIA}}
\author{Noe Demazure}
\affiliation{Lab. de physique ENS Lyon, F-69364 CEDEX 07, France }

\author{V\'{\i}ctor Gonz\'alez Sebasti\'an} 
\affiliation{ Dept. of Physics and Astronomy, Wayne State University, Detroit, Michigan 48201, USA}

\author{Felipe J. Llanes-Estrada}
\affiliation{Dept. F\'{\i}sica Te\'orica \& IPARCOS, Univ. Complutense de Madrid, 28040 Spain}

\date{\today}
\begin{abstract}
We present a computational investigation of a problem of hadron collisions from recent years, that of baryon anticorrelations. This is an experimental dearth of baryons near other baryons in phase space, not seen upon examining numerical Monte Carlo simulations.
We have addressed one of the best known Monte Carlo codes, {\tt PYTHIA}, to see what baryon (anti)correlations it produces, where they are originated at the string-fragmentation level in the underlying Lund model, and what simple modifications could lead to better agreement with data.\\
We propose two ad-hoc alterations of the fragmentation code, a ``one-baryon'' and an ``always-baryon'' policies 
that qualitatively reproduce the data behaviour, i.e anticorrelation,
and suggest that lacking Pauli-principle induced corrections at the quark level could be the culprit behind the current disagreement between computations and experiment.
\end{abstract}
\maketitle
%
%%%%%%%%%%%%%%%%%%%%%%%%%%%%%%%%%%%%%%%%%%%%%%%%%%%%%%%%%%%%%%%%%%%%%%%%%%%%%%%%%%%%%%%%%%%%%%%%%%%%

%%%%%%%%%%%%%%%%%%%%%%%%%%%%%%%%%%%%%%%%%%%%%%%%%%%%%%%%
\section{Introduction}
%%%%%%%%%%%%%%%%%%%%%%%%%%%%%%%%%%%%%%%%%%%%%%%%%%%%%%%%

%%%%%%%%%%%%%%%%%%%%%%%%%%%%%
\subsection{Baryon anticorrelations}
%%%%%%%%%%%%%%%%%%%%%%%%%%
The baryon anticorrelation problem was exposed in a series of works by the 
ALICE collaboration~\cite{Graczykowski:2014mta, ALICE:2016jjg}, investigations 
derived therefrom~\cite{Graczykowski:2021vki}, and further corroborated by the 
STAR collaboration~\cite{STAR:2019cqg}. Basically it is a severe disagreement about 
the probability of two baryons exiting a high-energy collision closely in 
phase-space (similar momenta). 
Computational simulations using event generators fail to see an anticorrelation
clearly present in the data. 
To understand it, let us first define the appropriate 
two-particle correlations that are measured.

Let us begin with the definition of the single and pair densities of species 
$\alpha$ and $\beta$ as
\begin{align}
\rho_1^{\alpha}(\eta_1,\varphi_1) 
  &\equiv  
    \frac{{\rm d}^2N_1^{\alpha}}{{\rm d}\eta_1{\rm d}\varphi_1}, \\ 
\rho_2^{\alpha\beta}(\eta_1,\varphi_1,\eta_2,\varphi_2) 
  &\equiv 
    \frac{{\rm d}^4N_2^{\alpha\beta}}{{\rm d}\eta_1{\rm d}\varphi_1{\rm d}\eta_2{\rm d}\varphi_2} 
\end{align}
where $N_1^{\alpha}$ represents the number of particles of species $\alpha$, $N_2^{\alpha\beta}$ 
represents the number of pairs of particles of species $\alpha$ and $\beta$, and $\eta_{\rm i}$ 
and $\varphi_{\rm i}$ represent the corresponding pseudorrapidity, $\eta$, and azimuthal
angle, $\varphi$, of the involved particle. Clearly the magnitude 
\begin{equation}
  \rho_2^{\alpha\beta}(\eta_1,\varphi_1,\eta_2,\varphi_2) -
  \rho_1^{\alpha}(\eta_1,\varphi_1)\rho_1^{\beta}(\eta_2,\varphi_2)
  \label{eq:corr}
\end{equation}
measures the degree of correlation between particles of species $\alpha$ in phase 
space bin $(\eta_1,\varphi_1)$ and particles of species $\beta$ in phase space bin
$(\eta_2,\varphi_2)$. Two-particle correlations are usually reported, normalizing
the Eq.~(\ref{eq:corr}) magnitude, as normalized second order cumulants, and in 
relative separation in pseudorapidity, $\Delta\eta = \eta_1-\eta_2$, and azimuth,
$\Delta\varphi=\varphi_1-\varphi_2$
\begin{equation}
  R_2^{\alpha\beta}(\Delta\eta,\Delta\varphi) = 
  \frac{\rho_2^{\alpha\beta}}{\rho_1^{\alpha}\rho_1^{\beta}} - 1\text{.}
  \label{eq:r2}
\end{equation}
Experimentally the expression in the denominator is usually obtained directly in 
$\Delta\eta$, $\Delta\varphi$ by using the mixed events technique which builds
the $\rho_2$ distribution considering one pair component from one event and the
second pair component from a different event. By using this technique, the 
expression~(\ref{eq:corr}) is identically zero, because particle from different events are not 
correlated, and
\begin{equation}
  [\rho_1\cdot \rho_1] (\Delta\eta,\Delta\varphi) = \rho_2^{\rm mixed} (\Delta\eta,\Delta\varphi)
\end{equation}
is obtained. Then, what is usually reported as measured two-particle correlation 
function is
\begin{equation}
    C^{\alpha\beta}(\Delta\eta,\Delta\varphi)
      = R_2^{\alpha\beta}(\Delta\eta,\Delta\varphi) + 1
      = \frac{S^{\alpha\beta}(\Delta\eta,\Delta\varphi)}{B^{\alpha\beta}(\Delta\eta,\Delta\varphi)}
  \label{eq:correlationfctn}
\end{equation}
where $S$ is the ratio of the distribution density in relative rapidity $\Delta\eta$ 
and azimutal angle $\Delta\varphi$ of presumably correlated pairs (because they are taken 
from the same event), to the total number of pairs,
\begin{equation}
    S(\Delta\eta,\Delta\varphi)=\frac{1}{N^{\rm signal}_{2}}\frac{{\rm d}^2N_{2}^{\rm signal}}
      {{\rm d}\Delta\eta \, {\rm d}\Delta\varphi}\ ,
\end{equation}
and $B$ is the equivalent distribution of uncorrelated pairs (stemming from separate 
events, or ``mixed'' in the field's jargon)
\begin{equation}\label{isoB}
    B(\Delta\eta,\Delta\varphi)=\frac{1}{N^{\rm mixed}_{2}}\frac{{\rm d}^2N^{\rm mixed}_{2}}
      {{\rm d}\Delta\eta \, {\rm d}\Delta\varphi}
\end{equation}
The mixed event distribution in Eq.~(\ref{isoB}) corrects for the effects of limited 
acceptance in pseudorapidity on $S$ but, apart from a factor scale, it does not affect 
its azimuthal shape.
For this reason, $S$ may be studied alone to investigate the behaviour in 
$\Delta\varphi$ of the correlation function $C$.

\begin{figure}[ht]
\includegraphics[width=0.3\textwidth]{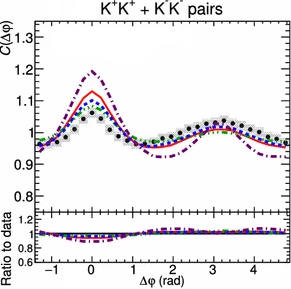}
\includegraphics[width=0.3\textwidth]{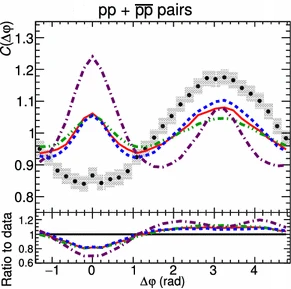}
\includegraphics[width=0.3\textwidth]{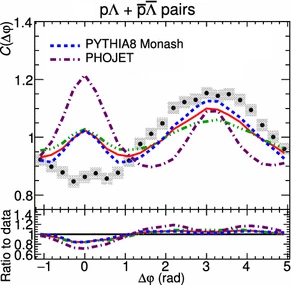}
\caption{{\bf Left:} Typical meson-meson correlation function from ALICE data~\cite{ALICE:2016jjg} 
as a function of $\Delta\varphi$, showing a large forward peak (two Kaons are more likely 
to be detected with $\Delta\varphi=\varphi_2-\varphi_1\simeq0$)
and a sizeable backward peak at $\Delta\varphi=\pi$. 
Simulation and experimental data are in reasonable qualitative agreement. 
{\bf Center and right:} Experimental baryon--baryon$+$anti-baryon--anti-baryion 
correlation functions become \emph{negative} (data, grey squares) when 
$\Delta\varphi\simeq 0$, showing a forward anticorrelation. 
Monte Carlo simulations seem unable to reproduce that behaviour.
Note that this happens not only for identical baryons such as $\rm pp$ (center) but 
also for distinguishable ones such as ${\rm p}\Lambda$ (right). 
Reproduced from~\cite{ALICE:2016jjg} by the ALICE collaboration under the terms 
of the Creative Commons License 4.0 ({\tt http://creativecommons.org/licenses/by/4.0/}); 
no changes have been effected.\label{problemdata} 
(The red and green lines correspond to {\tt PYTHIA} 6 with two Perugia tunes, 
and play no further role in our discussion). }
\end{figure}

Now let us describe the baryon anticorrelation problem. Contrasting experimental data 
with simulations from three tunes of the {\tt PYTHIA} Monte Carlo event generator
(two Perugia and the Monash one), the outcome was rather 
disappointing~\cite{Graczykowski:2014mta}.
The disagreement is not particularly due to {\tt PYTHIA} because its substitution by PHOJET, 
an alternative Monte Carlo event generator, did not bring improvement.
Figure~\ref{problemdata}, taken from \cite{ALICE:2016jjg}, shows the two-particle 
correlation functions.
Correlation functions are integrated over $\Delta\eta$ (and normalized), 
which is a more convenient to compare several results.

The OPAL collaboration discussed in~\cite{OPAL:2009hyp}
what was known on baryon correlations at the time of wrapping up the LEP experimental program.
Their conclusion was that, due to limited statistics and insensitivity to dynamics by lacking
rapidity in the analysis,  the correlations due to discrete quantum number
(flavor, charge, baryon number conservation were clearly visible, but not much beyond that: quark/fragmentation-level
correlations were not clearly established. The situation has now changed due to the impressive
statistics and tracking capabilities of more modern devices, and dynamical effects will need to be addressed.

%%%%%%%%%%%%%%%%%%%%%%%%%%%%%%%%%%%%%%%%%%%%%%%%%%%%%%%%%%%%%%%%%%%%%%
\subsection{Common use of the {\tt PYTHIA} software to predict correlations}
%%%%%%%%%%%%%%%%%%%%%%%%%%%%%%%%%%%%%%%%%%%%%%%%%%%%%%%%%%%%%%%%%%%%%%

{\tt PYTHIA} is a Monte Carlo event-generator software program, widely used to simulate 
possible outcomes of experiments (like collisions between two objects, be it 
fundamental particles or nuclei) and help understand detector responses.  
Through various physical model mechanisms, {\tt PYTHIA} evolves the initial input 
state into a number of output particles. 
These are listed by their flavor, spin and momentum in its original 
format described in \cite{Bierlich:2022pfr}.

Such simulation packages are popular because of their convenience to explore 
possible experimental outcomes when designing a measurement or during its 
later analysis. Moreover, {\tt PYTHIA} allows one to quickly test theoretical ideas, 
for example, adding some Beyond the Standard Model feature, to start assessing 
its impact on the data, even to guess physics at energies that current colliders 
cannot reach. 
Finally, finding a way to simulate a process can help understand it, 
so the shortcomings of {\tt PYTHIA} can sometimes reveal shortcomings in current 
knowledge.

{\tt PYTHIA} addresses those goals by quickly generating a great number of simulated events. 
The main requirements on the software are speed of execution and fidelity 
of the data produced at the end of the simulation compared to the data 
from experiment. 
On the contrary, fidelity to physical behaviour in the intermediates steps 
is secondary: the description of physical mechanisms used by {\tt PYTHIA} can 
be highly phenomenological, notwithstanding its inspiration in the ideas
of flux tubes forming among color sources and breaking in Quantum Chromodynamics~\cite{Greensite:2011zz}. 
It is therefore more descriptive than predictive, and can disagree with 
experiment when confronted with a new type of data. 
The phenomenological parametrizations deployed tend to use many parameters 
not set by any theory, but in order to fit the data. 
There are several possible ensembles of their values, called ``tunes''.

Returning to the left plot of Fig.~\ref{problemdata}  for meson-meson 
correlations, we see that the results of the {\tt PYTHIA} simulation are not 
far from experiment. 
The usual features are a strong correlation at small phase-space separation 
($\Delta \varphi \simeq 0 \simeq \Delta \eta$) for particles that exit 
the collision together, perhaps from the same jet; 
a positive correlation for back-to-back particles with opposite azimuthal 
angles  (a peak at $\Delta\varphi=\pi$ from two-jet events) 
and often (not visible in this flat depiction of the $\Delta\varphi$ 
dependence alone) a ridge-shaped correlation extending in $\Delta\eta$ 
often attributed to string-like flux tubes stretched by the separating nuclei. 

For baryon and anti-baryon correlations, shown in the center and right panels 
of  Fig.~\ref{problemdata}, however, the Monte Carlo simulation is unable 
to obtain a satisfactory description. 
The worst agreement is for baryon--baryon correlations, where {\tt PYTHIA} does not 
even produce the right qualitative behaviour. 
At around $\Delta\varphi=0$, all the tunes considered predict a peak instead of the 
salient depression (negative correlation, that is, anticorrelation) 
shown by the experiment. 
We would like to understand what modification of {\tt PYTHIA} would suppress that 
peak in baryon-baryon correlations and produce an anticorrelation instead.

Changing the ``tune'' (particular parameter set) of the simulation does not 
really improve the predicted correlations. 
Tunes can balance the influence of the different mechanisms inside {\tt PYTHIA}, 
but cannot change them drastically. 
Whatever reason for the failure of {\tt PYTHIA} in reproducing the data, it has to be 
deeper. 
Therefore we will limit ourselves to using the Monash tune in this work.

%%%%%%%%%%%%%%%%%%%%%%%%%%%%%%%%%%%%%%%%%%%%%%%%%%%%%%%%%%%%%%%%%%%%%%%%%%%%%%%%%%%%%%
\subsection{Recent attempts: the afterburners method}\label{subsec:afterburners}
%%%%%%%%%%%%%%%%%%%%%%%%%%%%%%%%%%%%%%%%%%%%%%%%%%%%%%%%%%%%%%%%%%%%%%%%%%%%%%%%%%%%%%

To correct the results of the simulation, a postprocessing procedure of the 
data has been proposed~\cite{Graczykowski:2021vki}.  
In the following, it will be named the ``afterburners'' procedure 
(in analogy to motor technology).  
This correction to the {\tt PYTHIA} output produces encouraging results for 
generic kinematics, but in this work we will only evaluate the very specific 
problem of proton-proton anticorrelation at $\Delta\varphi=0$, only one of 
several physical situations~\cite{Graczykowski:2021vki} but a problematic one: 
the correction to the $\rm pp$ correlation (see Fig.~2 in that article) 
is insufficient to account for the anticorrelation.

The idea behind this procedure is to write in addition a piece of Fortran code 
that from the output of {\tt PYTHIA} associates to each pair of particles a weight. 
The same-event correlation function $S$ is computed and binned as an histogram, 
which allows to weight each pair. 
In the simplest situation (no coupled channels, non identical particles), 
the weight is equal to the square modulus of the following wave function
\begin{eqnarray}
    &\psi({\bf k^*},{\bf r^*})=e^{i\arg\Gamma(1+i\eta)}\sqrt{A_C}\,\times\\
    &\qquad\qquad \left(1+\sum_{n=1}^\infty\frac{\zeta^n\prod_{m=0}^{n-1}
    (-i\eta+m)}{n!^2}+\frac{e^{i{\bf k^*}\cdot{\bf r^*}}\sqrt{A_C}
    \left(G_0(\rho,\eta)+iF_0(\rho,\eta)\right)
    /r^*}{f_0^{-1}+\frac{d_0{k^*}^2}{2}-ik^*A_C-
    \frac{2}{a}\left(\sum_{n=1}^\infty\frac{\eta^2}{n(n^2+\eta^2)}-
    \gamma-\ln|\eta|\right)}\right)
\end{eqnarray}
with
\begin{equation}
    \eta=\frac{1}{ak^*} \qquad \rho=k^*r^* \qquad \zeta={\bf k^*}{\bf r^*}+k^*r^* 
    \qquad A_C=\frac{2\pi\eta}{e^{2\pi\eta}-1}
\end{equation}
The variables ${\bf k^*}$ and ${\bf r^*}$ are the momentum and position 
vectors in the pair rest frame, $\gamma$ is Euler's constant and 
$f_0$, $d_0$ and $a$ are parameters given for each pair. 
Further, $F_0$ and $G_0$ are the regular and singular s-wave Coulomb functions. 
For further explanations, see \cite{Graczykowski:2021vki} and \cite{Lednicky:2005tb}.

The implementation of this final-state wavefunction allows to correct the 
correlation function $C$ to account for the antisymmetry upon exchanging the 
final-state baryons, unlike in the raw {\tt PYTHIA} output.
That computation of the weights is thus grounded both on quantum statistics 
and on final-state interactions due to Coulomb and nuclear forces. 
The exact formula giving the weight of a pair from the experiment, 
from the momentum and position  of each particle (classically treated by {\tt PYTHIA}, 
as explained below in section~\ref{sec:fragment}) is quite complex.

\begin{figure}[h]
    \centering\vspace{-0.5cm}
    \subfloat[]{\includegraphics[scale=0.35]{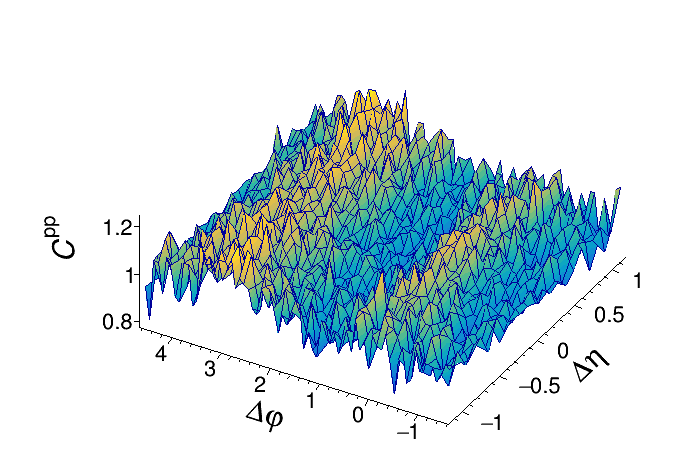}}
    \subfloat[]{\includegraphics[scale=0.35]{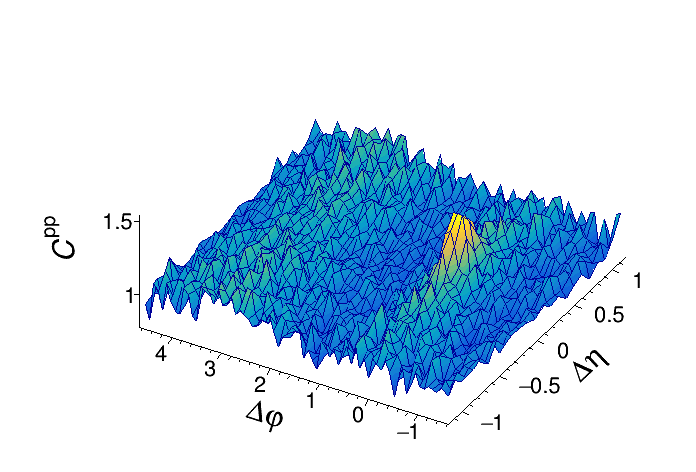}}
    \caption{Changes to the prediction of the proton-proton $R_{2}$ correlation 
    function induced by the ``afterburners'' correction. 
    (a) Proton-proton correlation predicted directly from the {\tt PYTHIA} Monte Carlo 
    program, without correction. 
    (b) Predicted proton-proton correlation corrected by the afterburners method. 
    Note that the positive
    correlation at the surroundings of $\Delta \varphi\simeq 0$ 
    has not been erased.}
    \label{Ledfig}
\end{figure}

We have attempted to redeploy that proposed method in our computations: 
the resulting modified correlation functions are shown in Fig.~\ref{Ledfig}. 
One can see at first glance that the data behaviour is not fully reproduced:  
the ridge at $\Delta\varphi=0$ present in the panel labeled as 
(a) is not totally suppressed in (b); 
and further, the correlation now strongly peaks at 
$(\Delta\eta,\Delta\varphi)=(0,0)$.

Though this method does not seem to alleviate the baryon-anticorrelation 
problem in our pedestrian simulation, 
the existence of this new peak is not completely problematic. 
The middle plot of Fig.~\ref{problemdata} shows a tiny peak in the middle 
of the anticorrelation valley (probably not a statistical blip, 
it is also clearly visible in three-dimensional renderings where 
it stands out for $\Delta \eta =0$. 
It appears that the afterburners correction might be predicting such peak. 
But (at least in our implementation) it does not produce the wide depression 
surrounding it.

Anyway, the afterburners procedure corrects the simulation output for all 
possible correlation functions: both $\rm pp$ and $\pi\pi$ are affected 
(in this case, the wavefunction is of course symmetrized).

Given this quick calculation, and the more extensive one~\cite{Graczykowski:2021vki} 
reported by the Warsaw Technical University investigators, 
we think  an additional way to correct the {\tt PYTHIA} output needs to be found.

%%%%%%%%%%%%%%%%%%%%%%%%%%%%%%%%%%%%%%
\section{Fragmentation in {\tt PYTHIA}} \label{sec:fragment}
%%%%%%%%%%%%%%%%%%%%%%%%%%%%%%%%%%%%%%
To motivate to the reader the modification that we are going to effect, we 
briefly expose our understanding of the 
{\tt PYTHIA} algorithms to produce final state hadrons. 
Then in the next section~\ref{sec:correlations} we will show the correlations 
that {\tt PYTHIA} produces before modification.

%%%%%%%%%%%%%%%%%%%%%%%%%%%%%%%%%%%%%%
\subsection{Simplified working method}
%%%%%%%%%%%%%%%%%%%%%%%%%%%%%%%%%%%%%%

A wise initial focus to minimize intervention in a time-tested code is to 
try to solve the problem at one single step of the algorithm. 
Since the afterburner postprocessing in subsection~\ref{subsec:afterburners} 
proved insufficient, this step would probably be at a previous level inside 
{\tt PYTHIA} 
(The earlier a modification is applied, the larger its effect upon cascading 
over the rest of the simulation.)

On the one hand, unstable-hadron decays cannot be the source of the problem, 
for they are simple processes in the code mimicking experimentally measured 
decays, so a disagreement with experiment is hard to expect.

On the other hand,  computed meson-meson correlations broadly agree with 
experiment hinting that we should modify only the baryon-production algorithm. 
To effect a difference between baryons and mesons supposes that hadrons 
have emerged from the colliding system at this step.

Additionally, QCD amplitudes are also well understood so the perturbative 
matrix elements affecting the initial stages of the process are unlikely to be 
the culprit.

So we adopt the point of view that if there is one single mechanism to be 
corrected in order to obtain the experimental behaviour of the correlation 
function $C$ for proton pairs in Fig.~\ref{problemdata}, it must occur
at the step of the formation of hadrons.

Because of the reported ${\rm p}\Lambda$ anticorrelation (right panel of 
Fig.~\ref{problemdata}), the wanted mechanism has to be generic, 
not tied to the nucleon.

We then proceed in the next subsection to summarize the 
way~\cite{Ferreres-Sole:2018vgo} {\tt PYTHIA} obtains prompt hadrons 
given initial partons.

%%%%%%%%%%%%%%%%%%%%%%%%%%%%%%%%%%%%%%%
\subsection{Lund string model}
%%%%%%%%%%%%%%%%%%%%%%%%%%%%%%%%%%%%%%%

The {\tt PYTHIA} level just before the formation of hadrons handles quarks and 
gluons organized into color singlets. Any interactions involving different 
such singlet clusters had been taken into account earlier during the parton 
level computations and can probably be ignored for hadronization. 
At the stage of hadron formation, the code takes as input a set of partons and 
returns hadrons, modelling low-energy QCD interactions. Due to its complexity 
and for speed of algorithm execution, {\tt PYTHIA} uses a phenomenological model 
called the Lund string model. 
Being far from perfect, this is convenient to treat hadron formation with 
sufficient richness of detail while maintaining acceptable turnout.

Summarizing the next few paragraphs, {\tt PYTHIA} views the input color singlet made 
of quarks and gluons as a string that fragments into pieces (the hadrons). 
The simplest possible string is made of one quark, one antiquark and no gluons. 
For brevity, we will describe the process of fragmentation only in that case. 
For a complete description, we refer to the 
literature~\cite{Ferreres-Sole:2018vgo}.

In the internal machinery of {\tt PYTHIA}, each particle is modelled semiclassically, 
with exactly defined position and momentum, at
odds with Heisenberg's uncertainty. 
The Lund model thus makes sense on average; this is convenient if a great number 
of fragmentations are performed, as is the regular usage of {\tt PYTHIA}, at the cost 
of microscopic interpretability.

A model string stretches between a quark and an antiquark of opposite color, 
called ``ends''; their flavor is not significant for the QCD interaction.  
Due to our understanding of QCD, the binding energy is taken as proportional 
to the distance that separates the ends.  

If a  quark-antiquark pair of the right color appeared at some point of the 
string (a tunnel-like effect), each parton of the pair could bind with an end 
to form its color singlet. 
Thus, the system would become two strings. 
If the total length of these two new strings was lower than the length of the 
original one, the situation would be energetically favorable.

The momentum of these partons emerging from the vacuum is chosen such that one 
of the daughter strings turns into a physical hadron of a certain flavour and 
invariant mass. This constitutes ``an iteration''. During the next iteration, 
the remaining daughter string becomes the mother one. This loop stops when all 
of the string's energy has been taken away by hadrons.

Instead of creating a $\rm q\bar{q}$ pair, {\tt PYTHIA} can also opt to produce a 
diquark-antidiquark one. 
{\tt PYTHIA} views the diquark made of two quarks as a single particle.
Such diquark can form a color singlet with a quark, just like an antiquark 
would. 
The only difference is that the hadron produced is a baryon instead of a 
meson\footnote{As it is, a diquark and an anti-diquark might try to form a 
hadron together. This case is handled in~\cite{Andersson:1984af}}.

The hadrons that have been produced by this process can be of any flavour, for 
example containing charm or other types of spatial or spin excitations. 
Most of them are going to decay in the next step of {\tt PYTHIA}, to finally give 
particles stable against strong-force mediated decays, such as pions, nucleons, 
kaons, Lambdas etc.

%%%%%%%%%%%%%%%%%%%%%%%%%%%%%%%%%%%%%%%%%%%%%%%%%%%%%%%%%%5
\section{Correlations among strings and among baryons in {\tt PYTHIA} fragmentation} \label{sec:correlations}
%%%%%%%%%%%%%%%%%%%%%%%%%%%%%%%%%%%%%%%%%%%%%%%%%%%%%%%%%%%

In section~\ref{sec:fragment} we described the passage from a string to prompt 
hadrons in terms of the color flow. 
Flavour is of lesser interest because at the end, most heavy-flavoured quarks 
later disappear following weak decays and are often not detected.

More determining is the momentum of the objects implied in the fragmentation. 
It so happens that the theoretical study of the hadrons' momentum distribution 
function is burdensome (so that how to apply some level of Pauli suppression 
depending on the momentum distributions is not obvious to us). 
Earlier work~\cite{Ferreres-Sole:2018vgo} addresses momentum distributions at 
the level of one iteration, but extracting the momentum distributions from the 
full fragmentation loop is not trivial. 

Because theoretical computation of the momentum distributions (both string 
input and hadron output to the fragmentation process) is unassailable,  
plotting the output data from simulations is the only way to understand 
their global behaviour.

%%%%%%%%%%%%%%%%%%%%%%%%%%%%%%%%%%%%%%%%%%%%%
\subsubsection{Hadron distributions} \label{subsec:baryondists}
%%%%%%%%%%%%%%%%%%%%%%%%%%%%%%%%%%%%%%%%%%%%%

To qualitatively understand the proton-proton two-particle correlation 
function~(\ref{eq:correlationfctn}) in $\Delta\varphi$, we put aside the 
denominator $B$, to focus on the numerator $S$, which is basically a two-particle
distribution affected by the limited longitudinal, $\eta$, acceptance which 
introduces the characteristic triangular shape in $\Delta\eta$. 
We expect the protons present in the final state to have been produced during 
the fragmentation of some string or by the decay of a heavier baryon, 
itself due to the fragmentation. 
Roughly, half of the baryons are supposed to decay into protons, half into 
neutrons and some into  $\Lambda$ hyperons. 
Thus, the distribution of protons in the final step should be comparable to the 
distribution of baryons just after the fragmentation step 
(respectively antiprotons to anti-baryons).

What we want to extract is the distribution $S$ of the pairs of baryons 
coming from the same event. 
Such same-event distribution will receive contributions from baryons ejected 
from the same string as well as baryons coming each from a separate string 
within the same event.
$S$ is then equal to the sum of the same-string baryon correlation plus 
the different-string baryon correlations.

\begin{figure}[h]
    \begin{center}
    \subfloat[$S_{\rm frag}$ Baryon-baryon from meson string]
    {\includegraphics[scale=0.25,keepaspectratio=true,clip=true,trim=30pt 0pt 20pt 40pt]
    {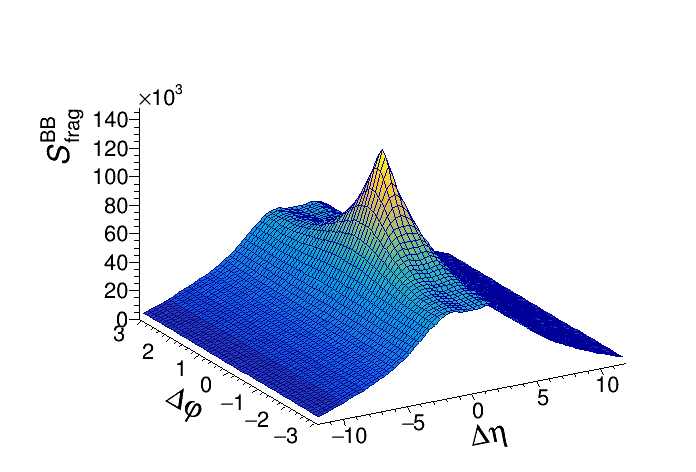}}
    \subfloat[$S_{\rm frag}$ Baryon-antibaryon from meson string]
    {\includegraphics[scale=0.25,keepaspectratio=true,clip=true,trim=30pt 0pt 20pt 40pt]
    {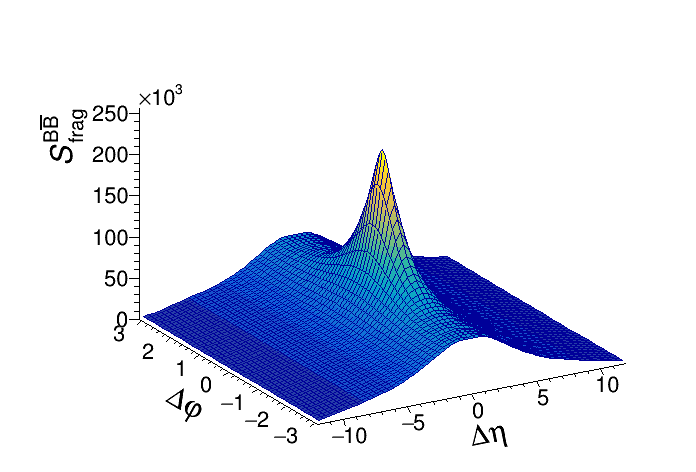}}
    \subfloat[$S_{\rm frag}$ Antibaryon-antibaryon from meson string]
    {\includegraphics[scale=0.25,keepaspectratio=true,clip=true,trim=30pt 0pt 20pt 40pt]
    {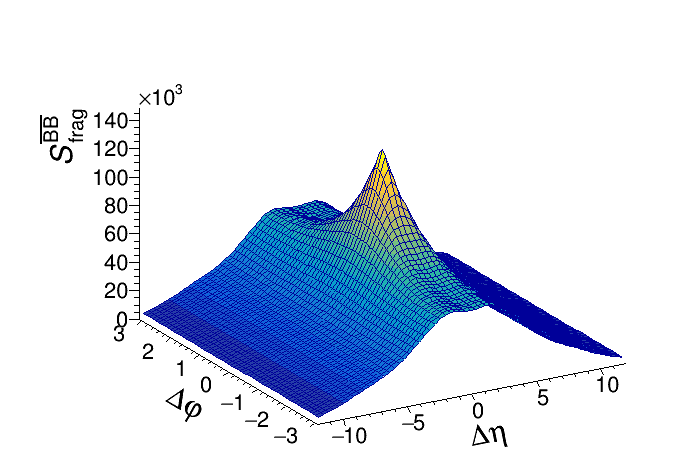}}\\
    \subfloat[$S_{\rm frag}$ Baryon-baryon from baryon string]
    {\includegraphics[scale=0.25,keepaspectratio=true,clip=true,trim=30pt 0pt 20pt 40pt]
    {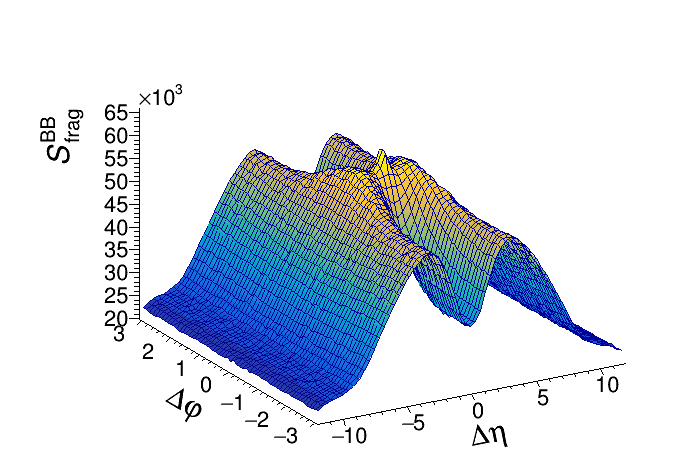}}
    \subfloat[$S_{\rm frag}$ Baryon-antibaryon from baryon string]
    {\includegraphics[scale=0.25,keepaspectratio=true,clip=true,trim=30pt 0pt 20pt 40pt]
    {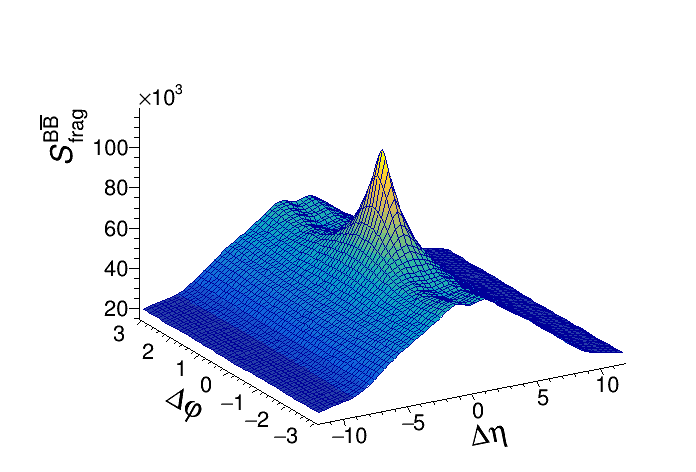}}
    \subfloat[$S_{\rm frag}$ Antibaryon-antibaryon from baryon string]
    {\includegraphics[scale=0.25,keepaspectratio=true,clip=true,trim=30pt 0pt 20pt 40pt]
    {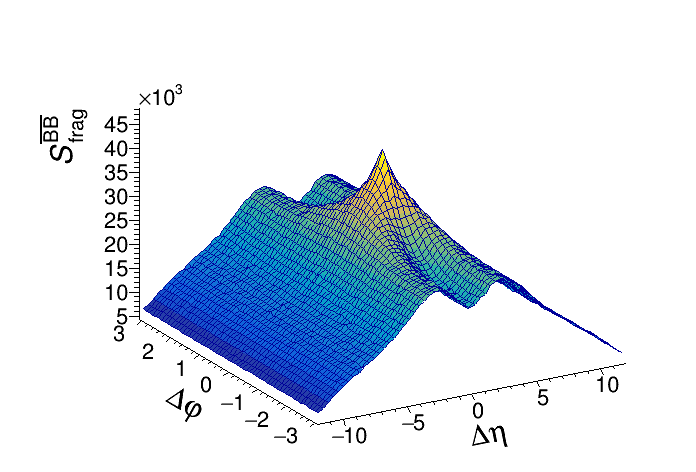}}
    \caption{Various two-particle baryon and/or antibaryon  correlations extracted at the level of strings in {\tt PYTHIA}. 
    By meson string we mean one with zero baryon number, while a baryon string contains baryon number one 
    (and thus, one net baryon is produced in the fragmentation). 
    By $S_{\rm frag}$ we mean that the two particles in each term of the correlation comes from the same string 
    within the same event.      
    \label{Str-BB}}
 \end{center}
\end{figure}

Because we seek to understand whether/why there are not suppressed correlations 
of baryons near to each other in phase space we will focus on  
\emph{baryons stemming from the same string}, denoting the corresponding 
correlation with $S_{\rm frag}$. 
For later comparability with data, the study is carried out in the lab frame.

The resulting numerical data is shown in Fig.~\ref{Str-BB}. 
Among the plots there, (a), (c), (d) and (f) will contribute to the 
proton-proton 
or antiproton-antiproton $S$ (if they were followed to the very end of the 
fragmentation). 

Independently of the baryon number combination, all pairs are positively 
correlated.
The only structure that seems to break the isotropy in $\Delta\varphi$ is the 
peak at around $(\Delta\eta, \Delta\varphi)=(0,0)$. 
Such peak accepts an easy interpretation: when a string fragments, baryons are 
preferentially produced along the same direction of the string, 
due to momentum conservation. 
Apart from that peak, (anti-)baryons from the pair are emitted with random 
$\varphi$ in the lab frame.

This strong forward peak at zero relative angle and pseudorapidity is precisely 
what seems to be suppressed in experimental data for $\rm BB$ and 
$\rm\bar{B}\bar{B}$ correlation functions, and we thus see the unwanted feature 
is already present at the level of individual string fragmentation.

The baryon string (bottom row panels) also produces two curious ridges most 
visible in panel (d) around $\Delta\eta= \pm 4$ for 13 TeV, 
which are outside ALICE's acceptance.

%%%%%%%%%%%%%%%%%%%%%%%%%%%%%%%%%%%%%%%%%%%
\subsubsection{Internal string distribution} \label{subsec:strings}
%%%%%%%%%%%%%%%%%%%%%%%%%%%%%%%%%%%%%%%%%%%

To ascertain whether the ``wrong'' baryon correlations in {\tt PYTHIA} are produced 
already at the string level, before the actual production of final-state 
particles, we have computed string-string correlations and report them in this 
section. 
As advanced in Fig.~\ref{Str-BB}, we distinguish two types of string, 
defined by their baryon number. 
If the initial two ends of a string are a quark and an anti-quark, the global 
baryon number is equal to 0, and we call it a meson string. 
If the ends hold a quark and a diquark, the baryon number is 1, and we call it 
a baryon string. 
Other cases happen to be so rare in {\tt PYTHIA} that they do not affect statistical 
distributions and are not further considered.

\begin{figure}[h]
    \centering
    \subfloat[$S_{\rm str}$ for meson string - meson string]
      {\includegraphics[scale=0.25,keepaspectratio=true,clip=true,trim=30pt 0pt 20pt 40pt]
      {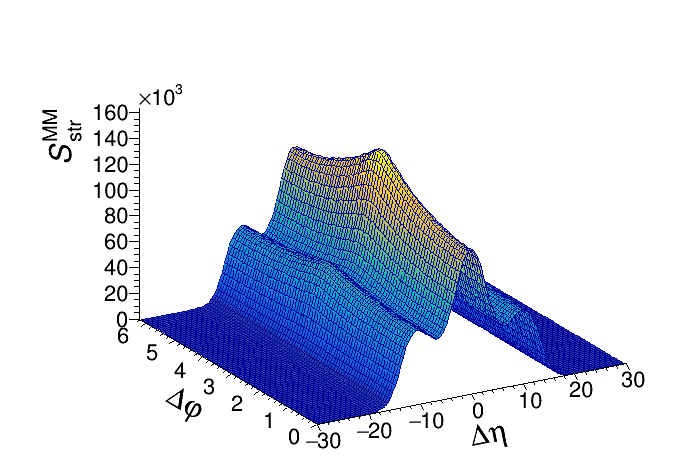}}
    \subfloat[$S_{\rm str}$ for meson string - baryon string]
      {\includegraphics[scale=0.25,keepaspectratio=true,clip=true,trim=30pt 0pt 20pt 40pt]
      {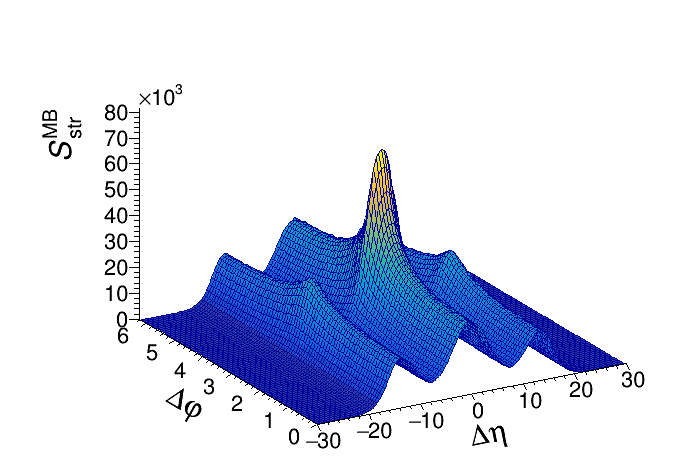}}
    \subfloat[$S_{\rm str}$ for baryon string - baryon string]
      {\includegraphics[scale=0.25,keepaspectratio=true,clip=true,trim=30pt 0pt 20pt 40pt]
      {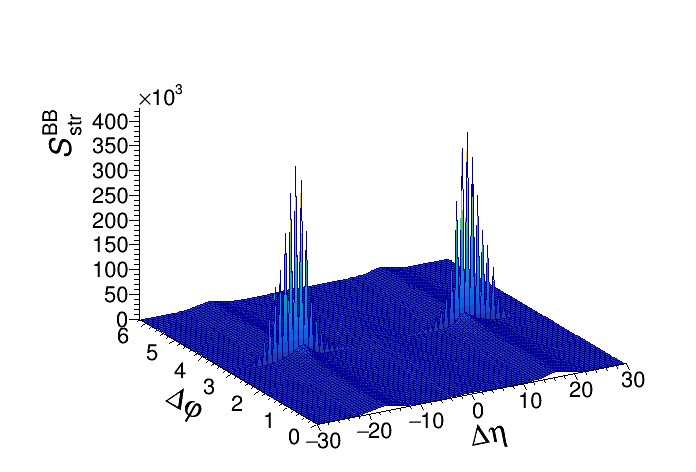}}
    \caption{String-string $S_{\rm str}$ correlations in $\rm pp$ collisions from 
    {\tt PYTHIA} showing clear  $\Delta\varphi=\pi$ peaks that would eventually 
    evolve to back-to-back jets. No kinematic cuts have been applied.
    As the strings carry a lot of energy, the pseudorapidity scale is not 
    comparable to measured hadrons.  
    Because high-$\eta$ hadrons escape ALICE's acceptance, 
    the contribution of the beams remnant from the original collided nuclei
    are erased in experimental data, but are clearly visible here in the uncut 
    simulation, as a pair of ridges at $|\Delta\eta|>10$.}
    \label{StrStr}
\end{figure}

Each string is viewed as a particle whose momentum (and thus its pseudorapidity 
$\eta$ and azimuth $\varphi$) is the sum over the string's constituents. 
The angular dependence of the two-string correlation~(\ref{eq:correlationfctn}) 
(unlike the absolute normalization) can be inferred from the same-event 
correlation $S_{\rm str}$, which we plot in Fig.~\ref{StrStr}.
In it we find that in the azimuthal dimension the distribution is mostly 
isotropic, except around $\Delta\varphi=\pi$, where we find a sizable peak. 
The relative importance of the isotropic and the peaking contributions varies 
with the participating strings. 
For meson-meson strings (a), the peak is tiny and the isotropic background 
offset is large. 
For baryon-baryon strings (c), it is the structure at $\Delta\varphi=\pi$  
(now separated in two peaks at large relative pseudorapidity) that dominates 
the graph, and the azimuthal isotropic offset is negligible in comparison. 
In (b), for meson-baryon strings, the situation is intermediate.

The pair of twin peaks at large relative pseudorapidity of order 10 or more in 
(c) is somewhat surprising.
It must be that the simulated collision between two protons produces, not only 
isotropically emitted baryons, but also two opposite beams. Whether a program 
artifact or an effect of the parent colliding protons remains to be seen, 
but in any case it is not currently comparable to the experimental data due 
to the limited rapidity coverage of the ALICE detector.

%%%%%%%%%%%%%%%%%%%%%%%%%%%%%%%%%
\subsubsection{Interpretation}\label{subsec:interpretation}
%%%%%%%%%%%%%%%%%%%%%%%%%%%%%%%%%

To summarize, the picture in $\Delta\varphi$ is the following: 
strings are quite uncorrelated except that they are likely to appear with 
opposite momenta (subsection~\ref{subsec:strings}). 

Baryons and anti-baryons from the same string (subsection~\ref{subsec:baryondists}) 
are likewise often randomly distributed, though also with a colinear component. 
A part of the explanation is that between every baryon pair
an antibaryon needs to be produced (to balance quark number) and this erases
azimuthal baryon-baryon correlations, for example. 
Baryons from the same string tend to be emitted in separate directions in the
string's reference frame, but the Lorentz boost given to the string 
upon passing to the laboratory frame appears to overcome this effect, so 
that baryons stemming from the same string are actually strongly collimated
and yield a forward peak in the final correlation.

The part of the hadrons that is uniformly distributed along $\Delta\varphi$ 
either from the same or from different strings are not supposed to provide any 
particular correlation in the final output. 
They only contribute to the isotropic offset, that influences the height of 
each peak or depth of each valley. 
As they do not change the qualitative behaviour, we do not discuss them any 
further in the following. 

The fraction of baryons and anti-baryons from the same string, colinearly 
emitted, will cause a peak around $\Delta\varphi=0$ in the final picture. 
The baryons and anti-baryons from different strings will contribute to 
the isotropic offset if their strings are uncorrelated and to a peak at 
around $\Delta\varphi=\pi$ if their strings are anti-colinear. 

Given that the number of uncorrelated strings is much larger than that of 
opposite-momentum strings, providing a large combinatorial background, 
it appears that the balance of the peaks at around $\Delta\varphi=0$, both for 
proton-proton and for proton-antiproton, are made of the correlations between 
the (anti-)baryons emitted by the same string~\footnote{This allows, 
from the number of baryons and anti-baryons fragmented per string, 
to give a bound to the number contributing to the final $\Delta\varphi=0$ peak 
in the two-particle correlation function. 
That is, if a string fragments into $N$ baryons and $N'$ anti-baryons, 
$N(N-1)$ pairs may contribute to the $\Delta\varphi=0$ peak in the final 
$\rm pp$ correlation function, $N'(N'-1)$ to the $\rm \bar{p}\bar{p}$ one and 
$NN'$ to the $\rm p\bar{p}$ one.
}.

This suggests to us that a modification to bring the Monte Carlo in better 
agreement with experimental data would be most efficient if applied to the 
production of baryons from the same string (that are naturally
more affected by Fermi statistics, even starting at the quark level).

%%%%%%%%%%%%%%%%%%%%%%%%%%%%%%%%%%%%%%%%%%%%%%%%%%%%%%%%%%%%%%%%%%%%%%%%%%%%%%%%%%%%%
\section{Suggested modifications to {\tt PYTHIA}
that seem to reproduce baryon anticorrelations as seen in data}
%%%%%%%%%%%%%%%%%%%%%%%%%%%%%%%%%%%%%%%%%%%%%%%%%%%%%%%%%%%%%%%%%%%%%%%%%%%%%%%%%%%%%

%%%%%%%%%%%%%%%%%%%%%%%%%%%%%%%%%%%%%%%%%%%%%%%%%%%%%%%%%%%%%%%%%%%%%%%%%%%%%%%%%%%%%
\subsection{First modification: one-baryon policy}
%%%%%%%%%%%%%%%%%%%%%%%%%%%%%%%%%%%%%%%%%%%%%%%%%%%%%%%%%%%%%%%%%%%%%%%%%%%%%%%%%%%%%

Following the explanation in subsection~\ref{subsec:interpretation}, the 
existence of the peak at around $\Delta\varphi=0$ in $\rm pp+\bar{p}\bar{p}$ 
correlation functions might be linked to strings giving birth to more than one 
hadron.

To test this idea, we completely forbid it in the program with a Von Neumann 
rejection step: anytime that a string fragments into more than one baryon 
(idem for anti-baryons), the fragmentation is recalculated.

%%%%%%%%%%%%%%%%%%%%%%%%%%%%%%%%%%%%%5
\subsubsection{Physical justification}
%%%%%%%%%%%%%%%%%%%%%%%%%%%%%%%%%%%%%%

One can see  this procedure as a way to impose the Pauli principle to baryons 
(which are fermions). 
The algorithm is not influenced by the nature of the baryons nor their spin. 
Instead, it could be influenced by the Fermi statistics of the quarks 
themselves. The similarity of the center ($\rm pp$) and right ($\rm p\Lambda$) 
plots in Fig.~\ref{problemdata}  hints that no difference should be made among 
species of baryons. 
At the baryon level this may sound strange since $\rm p$ and $\Lambda$ are 
distinguishable, but at the quark level, since both contain $\rm u$ and $\rm d$ 
valence quarks may be more understandable.

In the current {\tt PYTHIA} 8.3 version, Fermi-Dirac statistics does not seem to be 
implemented in any way. 
It would have been surprising that the absence of such an important principle 
was not falsified at some point by contrasting the generated data against 
experiment: it so has happened that multiple-baryon final states have not been 
thoroughly analyzed until this last decade. 

The appropriate theoretical procedure would be to anti-symmetrize an appropriate 
multi-fermion wavefunction or quantum amplitude, so we write down some formalism 
that could serve as illustration to fix ideas, although our computer-algorithm 
modifications will be much cruder.
A possible output of a fragmentation~\cite{Dowrick:1986ub} is associated with 
such an amplitude, derived from the wavefunctions of the involved entities 
through
\begin{eqnarray} 
    M(A\rightarrow BC)=\int d^3r\int d^3w \, \psi_B^*
      (\frac{1}{2}{\bf r}+{\bf w}) \psi_C^*
      (\frac{1}{2}{\bf r}-{\bf w}) {\bf \alpha}\cdot(i{\bf \nabla_B}+i{\bf \nabla_C}+{\bf q}) 
      \psi_A({\bf r}) e^{\frac{1}{2}i{\bf q}\cdot{\bf r}} \gamma^a_{bc}({\bf r},{\bf w})
\end{eqnarray} 
Were A, B and C are strings or hadrons, which are equivalent in this description. 
$\gamma^a_{bc}(r,w)$ is a function that can be computed with methods discussed 
in the literature~\cite{Dowrick:1986ub}.

It is easy to write a (perhaps long) expression for the semi-inclusive 
two-baryon production amplitude  of the form
\begin{eqnarray} 
    M(A\rightarrow B_1+B_2+X)=\int \; 
        \cdots \quad \psi_{B_1}^*(\cdots) \psi_{B_2}^*(\cdots) \quad \cdots
\end{eqnarray} 
That allows to anti-symmetrize the baryon part of the wavefunction as in an 
atomic Slater determinant
\begin{eqnarray} 
    \tilde{M}(A\rightarrow B_1+B_2+X)=\int \; 
        \cdots \quad \left(\psi_{B_1}^*\psi_{B_2}^*-\psi_{B_2}^*\psi_{B_1}^*\right) \quad \cdots
\end{eqnarray} 
and we expect $|\tilde{M}|<|M|$.

Taking the perhaps drastic but much simplifying approximation 
$\frac{|\tilde{M}|}{|M|} \rightarrow 0$ leads us to forbid the production of 
two baryons from the same string. 
This is the content of the first ``one-baryon only'' correction.

If one wanted to apply less dramatic simplification, one would have to 
redevelop parts of the Lund string model formalism to make it compatible with, 
for example, the formalism of~\cite{Dowrick:1986ub}.

%%%%%%%%%%%%%%%%%%%%%%%%
\subsubsection{Results}
%%%%%%%%%%%%%%%%%%%%%%%%

That simplest procedure of imposing a one-baryon per string policy can be very 
time consuming. 
Combining it with the afterburners correction worsens this and additionally 
leads to technical difficulties that makes it unpractical to gather sufficient 
Monte Carlo statistics in a limited university departmental cluster.

In the afterburners procedure, the lengthy code, together with the number of 
pairs increasing with the square of the number of events, can be a drawback. 
It brings about a rise of the whole time of execution, increases the frequency 
of software crashes and produces random pairs that occasionally give 
nonsense results (like negative weights) and must be discarded. 
If the afterburners method is used alone this is not dramatic and can be 
handled. But the difficulties compound if this addition to {\tt PYTHIA} is 
combined with another.

As it is, we renounce to deploy the two methods together. 
It would be interesting to find a time-economic implementation, but we have 
not managed to design one in a reasonably small time. 
Our one-baryon policy correction gives results that are sufficiently 
encouraging to confirm the relevance of our method and suggest that 
further work will be interesting.

\begin{figure}[h]
    \centering
    \subfloat[$\rm pp$ $C$ correlation with unmodified {\tt PYTHIA}]
    {\begin{minipage}[c][4.6cm][c]{7.3cm}\includegraphics[scale=0.3]{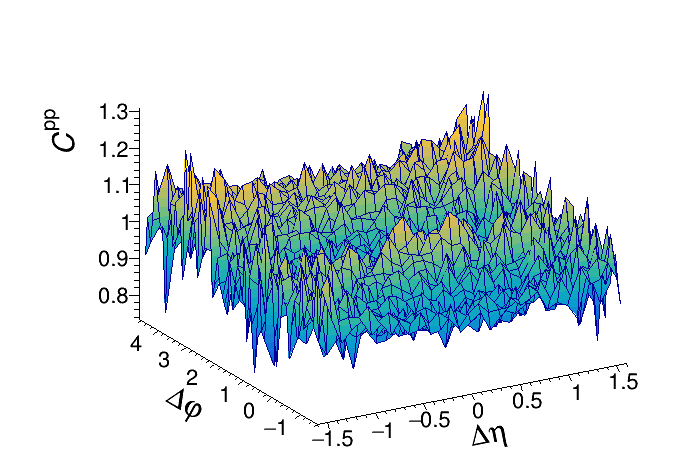}\end{minipage}}
    \subfloat[$\rm pp$ $C$ correlation, one-baryon per string policy]
    {\begin{minipage}[c][4.6cm][c]{7.3cm}\includegraphics[scale=0.3]{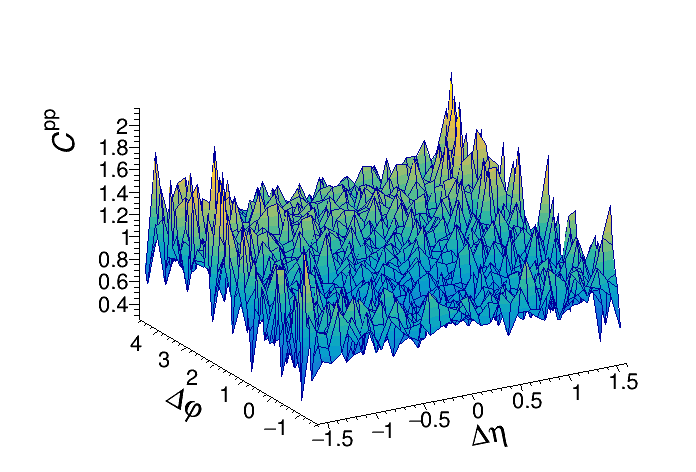}\end{minipage}}\\
    \subfloat[$\rm pp$ $C$ correlation, always-baryon policy]
    {\begin{minipage}[c][4.6cm][c]{7.3cm}\includegraphics[scale=0.3]{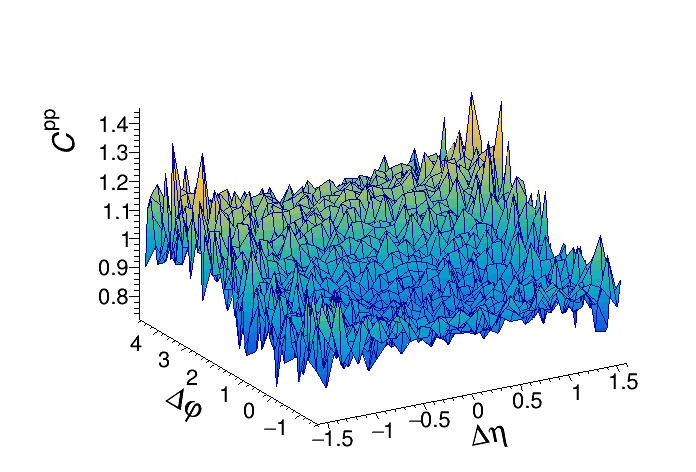}\end{minipage}}
    \subfloat[number of detectable (see \cite{ALICE:2016jjg} page 4) 
    protons produced in $5\times 10^7$ events.]
    {\hspace{0.5cm}\begin{minipage}[c][4.6cm][c]{7.3cm}
    \begin{tabular}{|r|c|c|c|}\hline
    Baryons produced ($10^6$) & $\rm p$ & $\rm pp$ & $\rm p\bar{p}$ \\\hline
    Uncorrected {\tt PYTHIA} & 5.44 & 1.52 & 2.61 \\\hline
    One-baryon policy & 1.55 & 0.14 & 0.44 \\\hline
    Always-baryon policy & 5.20 & 1.32 & 2.28 \\\hline
    \end{tabular}\\
    3rd and 4th columns : number of same-event pairs. Less pairs entail a 
    noisier figure.
    \end{minipage}}
    \caption{$\rm pp$ correlation function $C$ in pp collisions from 
    standard {\tt PYTHIA} 8 (top left) shows clear forward and backward ridges for 
    $\Delta\varphi=0,\pi$ and all relative pseudorapidities. 
    The one-baryon policy modification (top right) forbids two baryons from 
    the same string, erasing the forward correlation but depressing the total 
    number of baryons. 
    The additional modification of the always-baryon policy (lower left)
    solves this problem by increasing the overall number of baryons. 
    The backward correlation is still there; the forward correlation has now 
    been turned into an anticorrelation; and the number of baryons produced is 
    more adequate.}
    \label{Pauli_res}
\end{figure}

Figure~\ref{Pauli_res} (right top plot marked as b) shows that the first 
demand on the output is fulfilled: the peak at $\Delta\varphi=0$ has been 
suppressed by imposing to each string the one-baryon policy. 
Indeed, that peak seems to be produced by baryons coming from the same string. 
Still, this policy has a major side effect, which is to strongly reduce 
the number of protons in the final state (and also of $\Lambda$ hyperons 
and generically all baryon species). 
This was predictable because the method flatly reduces the average number of 
baryons produced by each string.

%%%%%%%%%%%%%%%%%%%%%%%%%%%%%%%%%%%%%%%%%%%%%%%%%%%%%%%%%%%%%%%%%%%%
\subsection{Second, balancing correction: no baryonless fragmentation or
``always baryons'' policy.}
%%%%%%%%%%%%%%%%%%%%%%%%%%%%%%%%%%%%%%%%%%%%%%%%%%%%%%%%%%%%%%%%%%%

To repurpose {\tt PYTHIA} to correct for the dearth of protons in the final state,
following the one-baryon policy, a second correction is tried. 

Recognizing that the overall baryon number has been reduced because each 
fragmentation is now constrained to produce one baryon at most, 
a simple way to rebalance it is to force each meson string to fragment into 
exactly one baryon and one anti-baryon (two baryons and one anti-baryon for 
baryon strings).

This second ``always baryons'' correction is a purely practical expedient that 
no physical explanation seems to justify within the string model. 
And yet, the results are satisfactory. 
Figure~\ref{Pauli_res} (lower-left plot marked c) shows that the relative 
features (ridges, valleys, absence of peaks) are equivalent to those of the 
one-baryon policy (upper right plot marked b), and it is only the overall 
number of baryons and anti-baryons which is increased. 
Any small remaining differences can probably be solved by a change of 
{\tt PYTHIA} tune: indeed, each change to {\tt PYTHIA} must induce a new calibration of 
its tune, but we leave this for future investigators and comment on it 
only briefly.

Among the parameters to be tuned, one  should be carefully  considered: 
$P_{Q\to QQ}$, the probability to engender diquark/anti-diquark pairs 
emerging from the vacuum during the string fragmentation, instead of 
quark/anti-quark ones. 
Changing this relative ratio is not independent of restraining the number of 
baryons emitted during the fragmentation. 
With our second always-baryons modification, this number  of baryons is 
imposed so this parameter has no impact on the final output. 

Nevertheless, an unwise choice of this $P_{Q\to QQ}$ parameter  has an 
unwelcome consequence. 
If one fragmentation iteration is unlikely to give exactly one baryon, 
it will be relaunched multiple times until it branches into a solution that 
satisfies the one-baryon policy. 
This may affect dramatically the length of execution of each fragmentation 
program loop.

But if only the first ``one-baryon policy'' correction is imposed, 
during fragmentation of meson strings 
the number of baryon--anti-baryon pairs produced can be either one or none. 
and that control parameter affects the relative probability of the two outcomes, 
becoming relevant to the output and not
only to the execution time.

Choosing a high value for this $P_{Q\to QQ}$ parameter can be viewed as getting 
closer to the second correction ``always-baryon policy'' even without 
imposing it. But increasing it causes the mean number of baryons in a 
string-fragmentation iteration to rise, and with it the possibility of 
discarding the fragmentation and having to rerun it due to exceeding 
the one-baryon policy. 
The effect on the global execution time could then be dramatic. 
Hence, the ``always-baryon policy'' can be viewed as an effective way of 
achieving the same effect without having to increase this parameter nor 
pay the extra execution time.

On the downside, this always-baryon policy is far from perfect from a 
computing implementation point of view. 
Compared to the one-baryon policy alone, it makes the code slower and more 
likely to crash. As it is, this implementation makes  a combination with the 
afterburners method impractical.

\begin{figure}[h]
    \centering
    \includegraphics[width=0.24\textwidth,keepaspectratio=true,clip=true,trim=40pt 0pt 50pt 40pt]
    {FIGS.DIR/OB_PrPPrP_C.png}
    \includegraphics[width=0.24\textwidth,keepaspectratio=true,clip=true,trim=40pt 0pt 50pt 40pt]
    {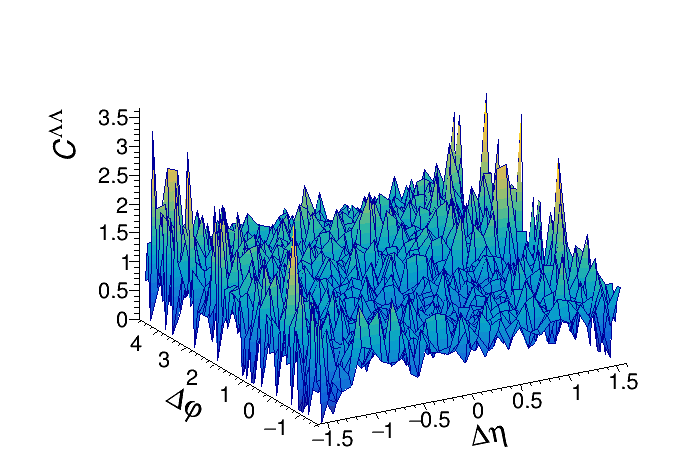}
    \includegraphics[width=0.24\textwidth,keepaspectratio=true,clip=true,trim=40pt 0pt 50pt 40pt]
    {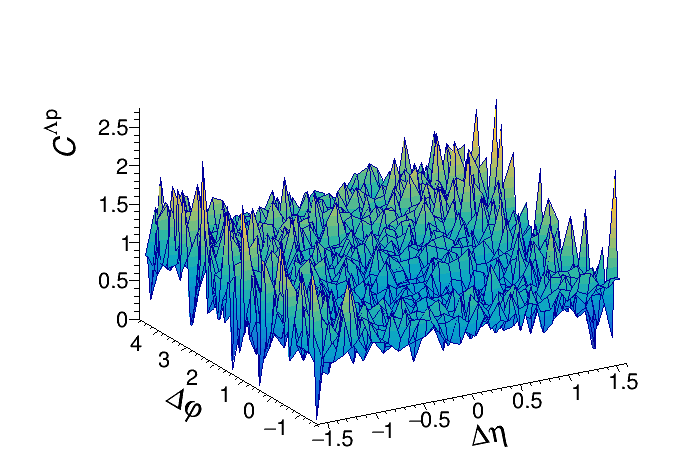}
    \includegraphics[width=0.24\textwidth,keepaspectratio=true,clip=true,trim=40pt 0pt 50pt 40pt]
    {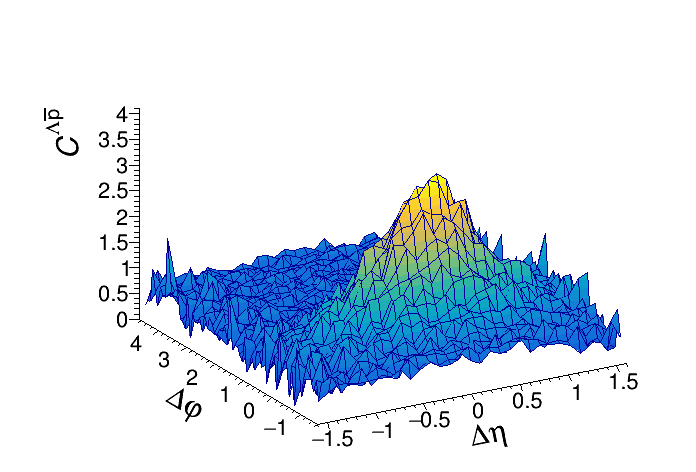}\\
    \includegraphics[width=0.24\textwidth,keepaspectratio=true,clip=true,trim=30pt 0pt 50pt 30pt]
    {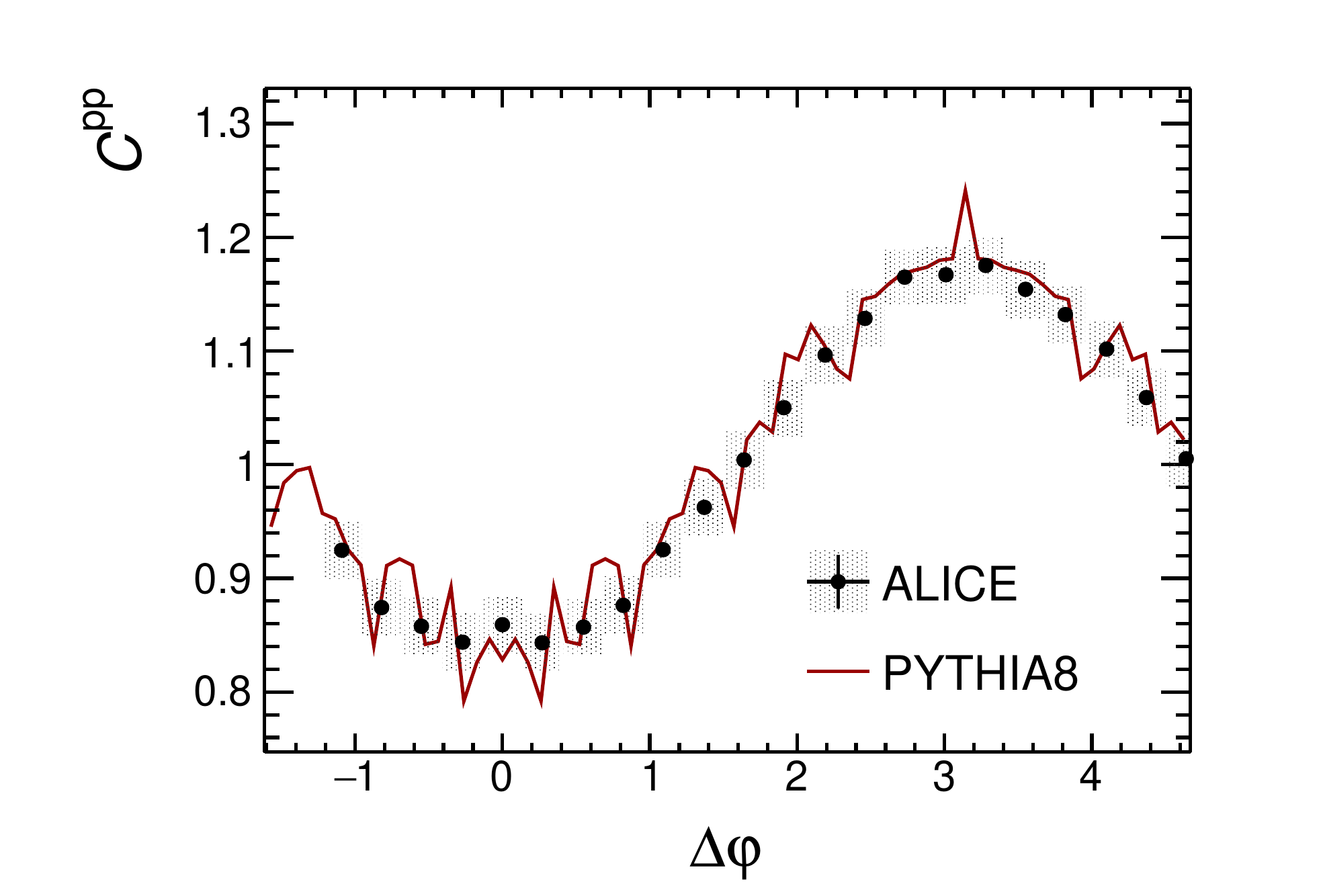}
    \includegraphics[width=0.24\textwidth,keepaspectratio=true,clip=true,trim=30pt 0pt 50pt 30pt]
    {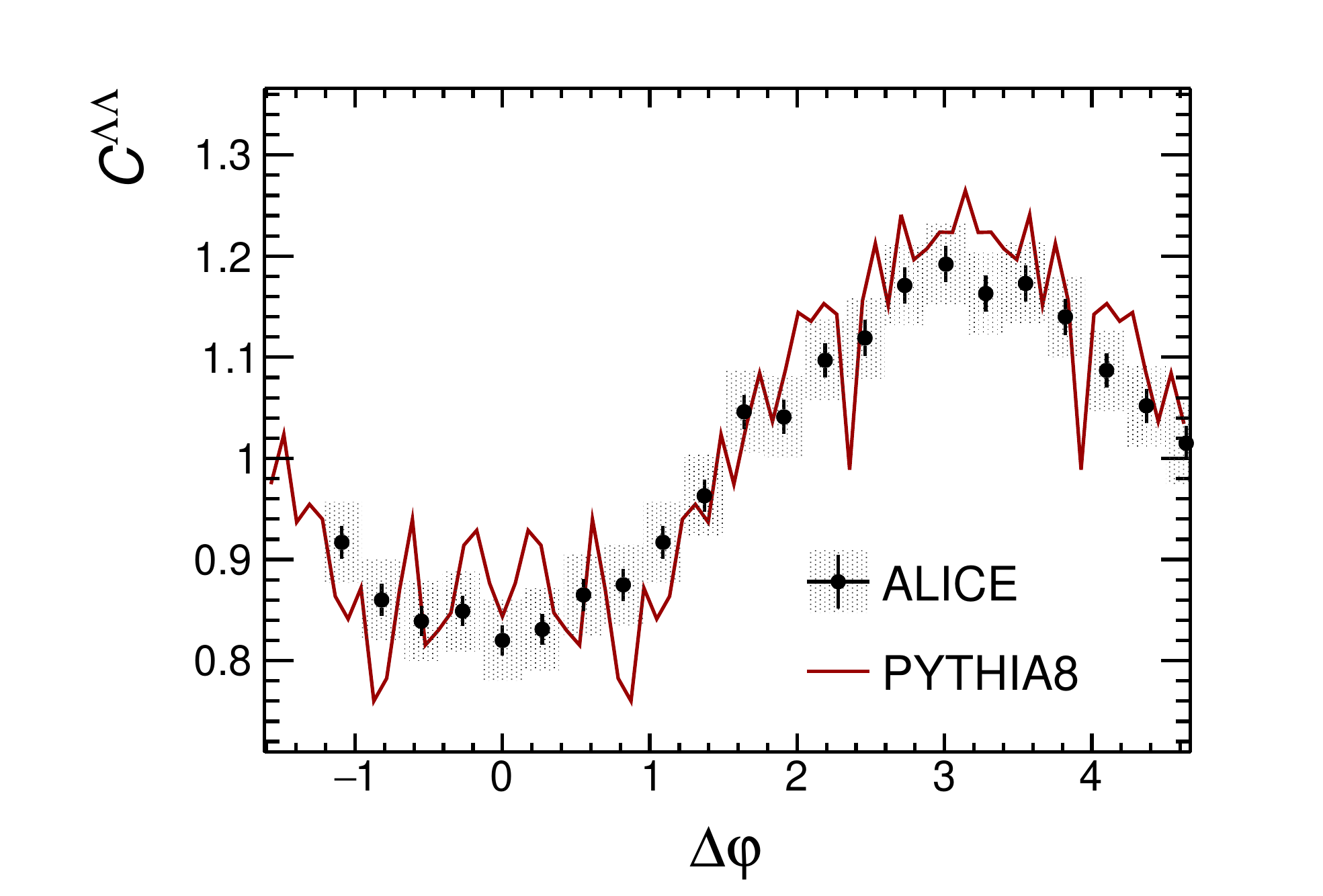}
    \includegraphics[width=0.24\textwidth,keepaspectratio=true,clip=true,trim=30pt 0pt 50pt 30pt]
    {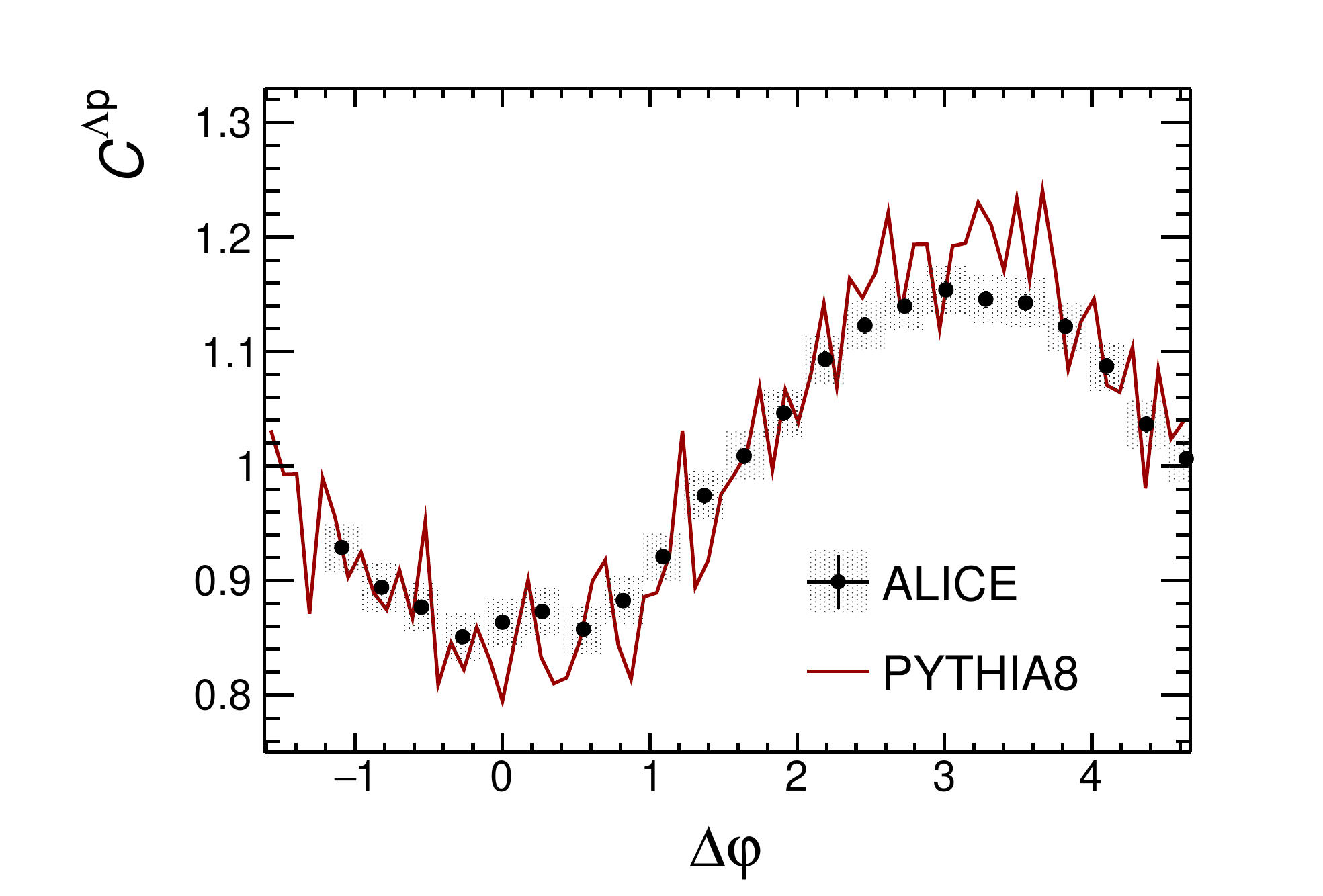}
    \includegraphics[width=0.24\textwidth,keepaspectratio=true,clip=true,trim=30pt 0pt 50pt 30pt]
    {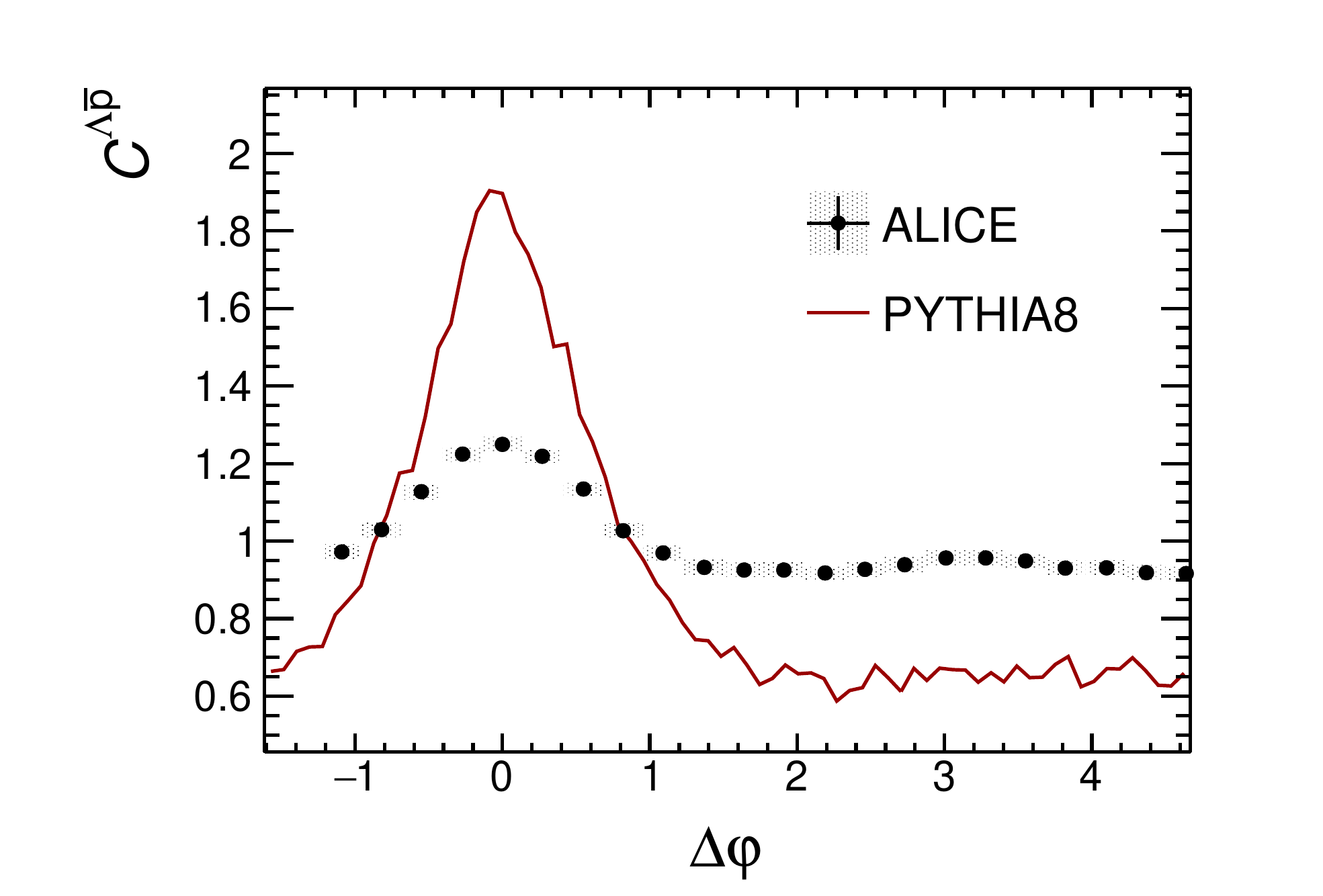}
    
    \caption{Correlation functions $C$ (top row) 
    for ${\rm pp}$, $\Lambda\Lambda$, $\Lambda\rm p$, 
    and $\Lambda\rm\bar p$ in pp {\tt PYTHIA}8 collisions applying the always-baryon policy. Azimuthal 
    projection (bottom row) of the correlation function $C$ for ${\rm pp}$, $\Lambda\Lambda$, 
    $\Lambda\rm p$, and $\Lambda\rm\bar p$ compared to ALICE data~\cite{ALICE:2016jjg}.
    }
    \label{BarABar_res}
\end{figure}

For completeness Fig.~\ref{BarABar_res} shows, top row, the correlation function $C$ from {\tt PYTHIA}8 
pp collisions, after applying the always-baryon policy, for ${\rm pp}$, $\Lambda\Lambda$, $\Lambda\rm p$, 
and $\Lambda\rm\bar p$ particle pairs. Bottom row shows the correlation function azimuthal projections 
compared to ALICE data~\cite{ALICE:2016jjg}.
Observe the excellent agreement between data and simulation for baryon-baryon correlations where the
anticorrelation observed in data is perfectly reproduced by the simulation. 
The strong discrepancy in the baryon-antibaryon correlation function compared to data shown in 
Fig.~\ref{BarABar_res} is not produced by the application of the always-baryon policy.
The discrepancy is already observed from {\tt PYTHIA}8 without introducing any modification as was also 
realized by the ALICE collaboration~\cite{ALICE:2016jjg}.

%%%%%%%%%%%%%%%%%%%%%%%%%%%%%%%%%%
\section{Conclusion and perspectives}
%%%%%%%%%%%%%%%%%%%%%%%%%%%%%%%%%%

We have examined the problem of baryon anticorrelations, a severe current 
disagreement between experimental data from the ALICE collaboration (lately also
from the STAR collaboration) and theoretical Monte Carlo simulations of hadron 
collisions. 
Such anticorrelations are typical of lower-energy nuclear physics, for example 
in two-proton radiactivity (see~\cite{Zhou:2022yzf}, right panel of Fig.~18, 
that shows a decrease at $\cos\theta_Y=-1$ when the protons are emitted back to 
back, respect to $\cos\theta_Y>-1$, an effect ascribed to Coulomb repulsion and 
the Pauli principle) but are now also routinely obtained in high-energy 
collisions, due to the improved particle identification; 
they should, and are not, reproduced by current event generators.

We believe that the ``fault'' lies on the theory side and that we have made a 
contribution to the discussion by exposing that modification of the Monte Carlo 
event generator  package {\tt PYTHIA} 8 at the fragmentation level can qualitatively 
account for anticorrelation. 
In our not very refined approach, this has been achieved by means of a 
``one-baryon policy'' applied to each string.
Because, with standard {\tt PYTHIA} parameters, this depresses the baryon to meson 
ratio (the total number of baryons produced is naturally smaller), a temporary 
correcting ``all-baryon policy'' has also been adopted, so that in our 
implementation given in the appendix, each string fragments into exactly one 
baryon (plus an unspecified number of mesons). 

If this problem was to be reassessed in future work, one could try to instead 
of suppress baryons from the same string, 
give them a natal kick to separate them (or more selectively suppress them 
depending on their produced momentum).
Another interesting theoretical extension would entail an extended probability 
distribution along the string, in order to mimic baryon or quark wave functions 
therealong.

Implementing a mechanism to take into account Fermi statistics in some way in 
{\tt PYTHIA} is really difficult if we want to keep its computer efficiency, 
reasonably fit the data and  give a convincing physical explanation. 
This is largely due to the fact that the Lund string model is mostly classical 
with point like particles while Fermi statistics is a quantum phenomenon based 
on wavefunctions. These two description are both essential, the first one for 
speed of execution and the second one for physical meaning. 
Thus, for the problem to be solved in a cleaner way, a translation between the 
two formalisms is necessary. 
For this, one must perhaps wait for a deeper understanding of the soft QCD 
processes that string fragmentation approximates. 
Conversely, finding a convenient procedure explaining major data features can 
give hints towards understanding of those deeper processes.

The LEP baryon correlation data could be reasonably fit by {\tt PYTHIA} as is, given that the color string did
form linking a back-to-back primary quark-antiquark pair;  this means that baryons from the same string did not 
form positive correlations near  $\Delta\eta \simeq 0 \simeq \Delta \varphi$ in OPAL data, 
as they were somewhat randomized, with the string frame not too far from the laboratory frame.

At the LHC strings are however formed at various rapidities and azimuths, with a natal Lorentz boost. 
Because of that string boost, two baryons formed from the same string will create that positive correlation in the laboratory frame.  Therefore, to avoid it and bring about the anticorrelation seen in the data, two-baryon production from the same
string should be suppressed: our way of achieving it is the very rough pair of policies (one-baryon and all-baryon)
that certainly need to be improved in future work.

%%%%%%%%%%%%%%%%%%%%%%%%%%%%%%%%%%%%%%%%%%%%%%%%%%%%%%%%%%%%%%%%%%%%%%%%%%%%%%%%%%%%%%%%%%%%%%%%%%%%
\section*{Acknowledgments}
%%%%%%%%%%%%%%%%%%%%%%%%%%%%%%%%%%%%%%%%%%%%%%%%%%%%%%%%%%%%%%%%%%%%%%%%%%%%%%%%%%%%%%%%%%%%%%%%%%%%
Noe Demazure thanks the members of the Theoretical Physics Department/IPARCOS institute at UCM for hosting him 
during this investigation, as well as the \'Ecole Normale Sup\'erieure Paris-Saclay for its financial support. 
The authors thank Claude Pruneau from Wayne State University for providing the WAC software package to represent 
the raw {\tt PYTHIA} output. They also thank Torbj\"orn Sj\"ostrand for useful comments on a first draft of the manuscript.\\
Work partially supported by the EU under grant 824093 (STRONG2020); by the US DOE under grant DE-
FG02-92ER40713; spanish MICINN under PID2019-108655GB-I00, PID2019-106080GB-C21; 
Univ. Complutense de Madrid under research group 910309 and the IPARCOS institute.

%%%%%%%%%%%%%%%%%%%%%%%%%%%%%%%%%%%%%%%%%%%%%%%%%%%%%%%%%%%%%%%%%%%%%%%%%%%%%%%%%%%%%%%%%%%%%%%%%%%%

\newpage

%%%%%%%%%%%%%%%%%%%%%%%%%%%%%%%%%%%%%%%%%%%%%%%%%%%
\newpage
\newgeometry{left=2cm,right=2cm,top=2.5cm,bottom=2.5cm}
\appendix
\section{Specific modification to {\tt PYTHIA}}
%%%%%%%%%%%%%%%%%%%%%%%%%%%%%%%%%%%%%%%%%%%%%%%%%%%
The most important modifications that we have described in the article are specified here.
We act on line {\tt{663}} of the {\tt PYTHIA} 8 version 307 package file

\hspace{0.5in}{\tt{/src/StringFragmentation.cc}}, 

\noindent in the function 

\hspace{0.5in}{\tt StringFragmentation::fragment}. 

\noindent The only other file that needs to be modified is 

\hspace{0.5in}{\tt{/src/{\tt PYTHIA}.cc}}, 

\noindent to adapt the validity checks accordingly.

We here provide the code of {\tt{StringFragmentation::fragment}} with the modifications highlighted.

\definecolor{coldef}{rgb}{1,1,1}
\definecolor{colcode}{rgb}{1,0.9,0.5}
\lstset {language=C++}
\begin{lstlisting}

// Perform the fragmentation.

bool StringFragmentation::fragment( int iSub, ColConfig& colConfig,
  Event& event) {
  
\end{lstlisting}\lstset{backgroundcolor=\color{colcode}}\begin{lstlisting}
  bool redo = settingsPtr->flag("Main:spareFlag1"); // Enable the baryon policy
  bool tolerance = settingsPtr->flag("Main:spareFlag2"); // can avoid crashes
  int nmin = settingsPtr->mode("Main:spareMode1");
  int nmax = settingsPtr->mode("Main:spareMode2");
  // minimal and maximal number of baryons per fragmentation
  if (redo && (nmin>=nmax || nmax<1)) cout << "bad choice of nmin, nmax" << '\n';

\end{lstlisting}\lstset{backgroundcolor=\color{coldef}}
\vspace{0.2cm}

(These flags and modes are the interface with the user to configure the hadron policy.)
\vspace{0.5cm}

\begin{lstlisting}

  // Find partons and their total four-momentum.
  iParton            = colConfig[iSub].iParton;
  iPos               = iParton[0];
  if (iPos < 0) iPos = iParton[1];
  int idPos          = event[iPos].id();
  iNeg               = iParton.back();
  int idNeg          = event[iNeg].id();
  pSum               = colConfig[iSub].pSum;
  
\end{lstlisting}\lstset{backgroundcolor=\color{colcode}}\begin{lstlisting}
  //Add the string to the event list.
  int Nq=0;
  for (int i:iParton){
    if (i>0) {
      int idp=event[i].id();
      if (idp<0 && idp>-10) Nq-=1;
      if (idp<10 && idp>0) Nq+=1;
      if (idp<-1000 && idp>-6000) Nq-=2;
      if (idp<6000 && idp>1000) Nq+=2;
    }
  }
  int B=Nq/3;
  if (abs(B)<2) event.append(52+B,201,iPos,iNeg,0,0,0,0,pSum,pSum.mCalc());
\end{lstlisting}\lstset{backgroundcolor=\color{coldef}}
\vspace{0.2cm}

(This part makes {\tt PYTHIA} return the strings together with the remaining intermediates particles of the event. 
It is needed to plot Figs.~\ref{Str-BB} and \ref{StrStr}.)

\vspace{0.5cm}
\begin{lstlisting}

  // Rapidity pairs [yMin, yMax] of string piece ends.
  vector< vector< pair<double,double> > > rapPairs = colConfig.rapPairs;

  // Reset the local event record and vertex arrays.
  hadrons.clear();
  stringVertices.clear();
  legMinVertices.clear();
  legMidVertices.clear();
  posEnd.hadSoFar = 0;
  negEnd.hadSoFar = 0;

  // For closed gluon string: pick first breakup region.
  isClosed = colConfig[iSub].isClosed;
  if (isClosed) iParton = findFirstRegion(iSub, colConfig, event);

  // For junction topology: fragment off two of the string legs.
  // Then iParton overwritten to remaining leg + leftover diquark.
  pJunctionHadrons = 0.;
  hasJunction = colConfig[iSub].hasJunction;
  if (hasJunction && !fragmentToJunction(event)) return false;
  int junctionHadrons = hadrons.size();
  if (hasJunction) {
    idPos  = event[ iParton[0] ].id();
    idNeg  = event.back().id();
    pSum  -= pJunctionHadrons;
  }

  // Set up kinematics of string evolution ( = motion).
  system.setUp(iParton, event);
  stopMassNow = stopMass;
  int nExtraJoin = 0;

  // Fallback loop, when joining in the middle fails.  Bailout if stuck.
  for ( int iTry = 0; ; ++iTry) {
    if (iTry > NTRYJOIN) {
      infoPtr->errorMsg("Error in StringFragmentation::fragment: "
        "stuck in joining");
      if (hasJunction) ++nExtraJoin;
      if (nExtraJoin > 0) event.popBack(nExtraJoin);
      return false;
    }

    // After several failed tries join some (extra) nearby partons.
    if (iTry == NTRYJOIN / 3) nExtraJoin = extraJoin( 2., event);
    if (iTry == 2 * NTRYJOIN / 3) nExtraJoin += extraJoin( 4., event);

    // After several failed tries gradually allow larger stop mass.
    if (iTry > NTRYJOIN - NSTOPMASS) stopMassNow
      *= (max( abs(posEnd.flavOld.id), abs(negEnd.flavOld.id)) < 4)
      ? FACSTOPMASS : FACSTOPMASS * FACSTOPMASS;

    // Set up flavours of two string ends, and reset other info.
    setStartEnds(idPos, idNeg, system);
    pRem = pSum;


    // Begin fragmentation loop, interleaved from the two ends.
    bool fromPos;

    // Variables used to help identifying baryons from junction splittings.
    bool usedPosJun = false, usedNegJun = false;

    // Keep track of the momentum of hadrons taken from left and right.
    Vec4 hadMomPos, hadMomNeg;

    // Inform the UserHooks about the string to he hadronised.
    if ( userHooksPtr && userHooksPtr->canChangeFragPar() )
      userHooksPtr->setStringEnds(&posEnd, &negEnd, iParton);

    for ( ; ; ) {

      // Take a step either from the positive or the negative end.
      fromPos           = (rndmPtr->flat() < 0.5);
      StringEnd& nowEnd = (fromPos) ? posEnd : negEnd;

      // Check how many nearby string pieces there are for the next hadron.
      double nNSP = (closePacking) ? nearStringPieces(nowEnd, rapPairs) : 0.;

      // The FlavourRope treatment changes the fragmentation parameters.
      if (flavRopePtr) {
        if (!flavRopePtr->doChangeFragPar(flavSelPtr, zSelPtr, pTSelPtr,
          (fromPos ? hadMomPos.m2Calc() : hadMomNeg.m2Calc()), iParton,
          (fromPos ? idPos : idNeg)) )
          infoPtr->errorMsg("Error in StringFragmentation::fragment: "
            "FlavourRope failed to change fragmentation parameters.");
      }

      // Possibility for a user to change the fragmentation parameters.
      if ( (userHooksPtr != 0) && userHooksPtr->canChangeFragPar() ) {
         if ( !userHooksPtr->doChangeFragPar( flavSelPtr, zSelPtr, pTSelPtr,
           (fromPos ? idPos : idNeg),
           (fromPos ? hadMomPos.m2Calc() : hadMomNeg.m2Calc()),
           iParton, &nowEnd) )
           infoPtr->errorMsg("Error in StringFragmentation::fragment: "
           "failed to change hadronisation parameters.");
      }

      // Check whether to use special hard diquark handling in beam remnant.
      bool forbidPopcornNow = false;
      if (forbidPopcorn && !hasJunction && (nowEnd.hadSoFar == 0)) {
        int iNow = (fromPos) ? iPos : iNeg;
        if (event[iNow].isDiquark()) {
          bool motherInBeam = (event[iNow].statusAbs() == 63);
          bool grannyInBeam = (event[event[iNow].mother1()].statusAbs() == 63);
          if (motherInBeam || grannyInBeam) forbidPopcornNow = true;
        }
      }

      // Construct trial hadron and check that energy remains.
      nowEnd.newHadron(nNSP, forbidPopcornNow);
      if ( energyUsedUp(fromPos) ) break;

      // Optionally allow a hard baryon fragmentation in beam remnant.
      bool useInputZ = false;
      double zUse    = 0.5;
      if (forbidPopcornNow && hardRemn) {
        useInputZ = true;
        zUse      = zSelPtr->zLund( aRemn, bRemn);
      }

      // Construct kinematics of the new hadron and store it.
      Vec4 pHad = nowEnd.kinematicsHadron(system, stringVertices,
        useInputZ, zUse);
      int statusHad = (fromPos) ? 83 : 84;
      nowEnd.hadSoFar += 1;

      // Change status code if hadron from junction.
      if (abs(nowEnd.idHad) > 1000 && abs(nowEnd.idHad) < 10000) {
        if (fromPos && event[iPos].statusAbs() == 74 && !usedPosJun)  {
          statusHad = 87;
          usedPosJun = true;
        }
        if (!fromPos && event[iNeg].statusAbs() == 74 && !usedNegJun)  {
          statusHad = 88;
          usedNegJun = true;
        }
        if (!fromPos && hasJunction && !usedNegJun) {
          statusHad = 88;
          usedNegJun = true;
        }
      }

      // Possibility for a user to veto the hadron production.
      if ( (userHooksPtr != 0) && userHooksPtr->canChangeFragPar() ) {
        // Provide full particle info for veto decision.
        if ( userHooksPtr->doVetoFragmentation( Particle( nowEnd.idHad,
          statusHad, iPos, iNeg, 0, 0, 0, 0, pHad, nowEnd.mHad), &nowEnd ) )
          continue;
      }

      // Bookkeeping of momentum taken away.
      if (fromPos) hadMomPos += pHad;
      else         hadMomNeg += pHad;

      // Append produced hadron.
      int colHadOld = nowEnd.colOld;
      int colHadNew = nowEnd.colNew;
      if ( !nowEnd.fromPos ) swap(colHadOld, colHadNew);
      hadrons.append( nowEnd.idHad, statusHad, iPos, iNeg,
        0, 0, colHadOld, colHadNew, pHad, nowEnd.mHad);
      if (pHad.e() < 0.) break;

      // Update string end and remaining momentum.
      nowEnd.update();
      pRem -= pHad;

    // End of fragmentation loop.
    }

    // Check how many nearby string pieces there are for the last hadron.
    double nNSP = (closePacking) ? nearStringPieces(
      ((rndmPtr->flat() < 0.5) ? posEnd : negEnd), rapPairs) : 0.;

    // When done, join in the middle. If this works, then really done.
    if ( finalTwo(fromPos, event, usedPosJun, usedNegJun, nNSP) ) {
    \end{lstlisting}\lstset{backgroundcolor=\color{colcode}}\begin{lstlisting}
      if (!redo) break;
      // redo if the number of baryons is not right
      if (iTry > 10 && tolerance) {
        infoPtr->errorMsg("Error in StringFragmentation::fragment: "
        "use tolerance");
        break;}
      else {
        int nBAB = 0; //number of baryons + antibaryons
        int idHad;
        for (int kount = 0; kount < hadrons.size(); ++kount) {
          idHad = hadrons[kount].id();
          if (idHad >  1000 && idHad <  10000) ++nBAB;
          if (idHad < -1000 && idHad > -10000) ++nBAB;
        }
        if (nBAB>=nmin && nBAB<=nmax) break;
      }
    }
\end{lstlisting}\lstset{backgroundcolor=\color{coldef}}
\vspace{0.2cm}

(If that condition is not fulfilled, the whole function is executed again. 
This corresponds to the one-baryon policy if $(n_{min},n_{max})=(0,2)$, 
and to the always-baryon policy if $(n_{min},n_{max})=(2,3)$.)
\vspace{0.5cm}

\begin{lstlisting}

    // Else remove produced particles (except from first two junction legs)
    // and start all over.
    int newHadrons = hadrons.size() - junctionHadrons;
    hadrons.popBack(newHadrons);
    stringVertices.clear();
    posEnd.hadSoFar = 0;
    negEnd.hadSoFar = 0;
  }

  // Junctions & extra joins: remove fictitious end, restore original partons.
  if (hasJunction) ++nExtraJoin;
  if (nExtraJoin > 0) {
    event.popBack(nExtraJoin);
    iParton = colConfig[iSub].iParton;
  }

  // Store the hadrons in the normal event record, ordered from one end.
  store(event);

  // Store hadron production space-time vertices.
  bool saneVertices = (setVertices) ? setHadronVertices( event) : true;

  // Done.
  return saneVertices;

}

\end{lstlisting}

\end{document}